\documentclass[11pt]{article}
\usepackage{times,aaspp4,astrobib,epsf}

\input{epsf}

\def\lsim{\hbox{ \rlap{\raise 0.425ex\hbox{$<$}}\lower 0.65ex\hbox{$\sim$} }}
\def\gsim{\hbox{ \rlap{\raise 0.425ex\hbox{$>$}}\lower 0.65ex\hbox{$\sim$} }}

\begin{document}

\title{ Line Caustic Crossing Microlensing and  Limb Darkening}
\author{Sun Hong Rhie and David P. Bennett}
\affil{Physics Department, University of Notre Dame, Notre Dame, IN 46556}

\centerline{(\today)}
\begin{abstract}

In a line caustic crossing  microlensing event, the caustic 
line moving across the surface of the source star provides  a direct
method to measure the integrated luminosity profile of the star. 
Combined with the enormous brightening at the caustic crossings, microlensing 
offers a promising tool for studying stellar luminosity profiles.
We derive the amplification behavior of the two extra images that become 
partial images conjoined across the critical curve at a line caustic crossing.
We identify the multiplicative factors that depend on 
the caustic crossing point and the relative size of the star, and the shape function
that depends on the stellar luminosity profile.   We examine  the analytic  
limb-darkening  models -- linear, square root, and square --  using the analytic 
form of the shape function.   We find that  the microlensing lightcurves must be
determined to an accuracy of better than $0.3$-$0.8\%$ 
in order to be able to determine the linear limb-darkening parameter $c_1$
with precision of $\delta c_1 = 0.1$.  This is similar to the accuracy level required
in eclipsing binaries as reported by Popper (1984).

\end{abstract}

\newpage

The caustic curve of a binary lens is where the number of images changes 
by two, and the two images that appear or disappear are extremely bright.
The caustic curve is a geometric curve with zero width, and
the differential limb darkening profile  of a lensed star (of a typical
size of $\sim 1-10 \mu$as) can be readily scanned during a caustic crossing.
The achromaticity of gravitational lensing and the large magnification of 
the two extra images help microlensing to be a rather clean  tool for 
studying the luminosity profiles of the lensed stars.

The microlensing method has several advantages: 
\ 1)~ The universality of gravitation guarantees that any star can be lensed,
and its limb darkening can be studied through microlensing .  That is not always 
the case in other methods.   For example, the lensed star of the 
recent line-caustic crossing binary microlensing event MACHO-98-SMC-1 
(\citeN{98smc1-mps}; \citeN{98smc1-joint} and references therein)  was a 
(spectroscopically) normal 
low metalicity A star \cite{98smc1-planet}.  Normal A stars have not been found
in short-period binary systems \cite{abt}, and their limb darkening
is unlikely to be studied through eclipsing. 
\ 2)~ The auxiliary object that accommodates high resolution scanning of the
lensed star  is a foreground (lensing) object that has no physical interaction
with the lensed star under study.  Combined with enormous brightening, the
physical independence  simplifies the interpretation of the event.
In an eclipsing binary, the very close interaction of the eclipsed star
with the companion (the auxiliary object) requires an elaborate modeling of the
binary system with many parameters \cite{popper} 
and makes it more difficult for the method to be robust.
\ 3)~  Microlensed stars are in the Galactic bulge ($\sim 8$~kpc) or in the  
Magellanic Clouds ($\sim 50$, $60$~kpc), and microlensing allows  
studies of the luminosity profiles of faint distant stars.

Resolving stars without auxiliary means has not been feasible technologically
except for the biggest stars in the sky: the Sun 
and more recently Betelgeuse (M2 supergiant) studied with HST \cite{betelgeuse}.  
Most of the studies of stellar luminosity profiles have been done with a handful
of detached eclipsing binaries.
In eclipsing binaries, the companion traverses in front of
the star and effectively  scans the luminosity profile of the eclipsed star
by absorbing the light. 
How well one can determine the profiles depends
on the well depth of the eclipse light curve, and it has been estimated
through numerical simulations  that 100 or more observations within minima
with a relatively small photometric error of $\sim 0.5\%$ are necessary 
to be able to determine the linear limb darkening parameter 
of the eclipsed star with the uncertainty of $\delta c_1 = 0.1$ or 
less \cite{popper}.
These simulations were done with simplified models,
and obtaining the precision of $\delta c_1 = 0.1$ may turn out  to be more 
demanding in practice. 

In a caustic crossing gravitational microlensing, the caustic curve 
traverses across the surface of the star and scans the integrated 
luminosity profiles of the star by partially magnifying the star.   
This  partial magnification is due to the  two extra images,
and the magnification behavior of them can be expressed as a simple analytic
function integrated over the luminosity profile.   We examine analytic models 
of the luminosity profiles and find that  relatively small photometric errors 
are required for the determination of the limb-darkening profiles:  ~In order
to be able to determine the linear limb darkening parameter $c_1$ with an 
uncertainty of $\delta c_1 = 0.1$ or less,  photometric accuracy of $0.3-0.8~\%$    
is required depending on the value of $c_1$ (see figure \ref{fig-delc} and 
\ref{fig-delc-norm}).

It has been a common practice to incorporate the effect of the luminosity profile
of the lensed star for the high resolution behavior of the light curve 
(for example, supernovae lensing by Schneider and Wagoner (1987)).
For stable stars, the linear limb-darkening law \cite{milne,gray} 
has been the default model for stellar luminosity profiles.
\begin{equation}
  {I(h)\over I_{\circ}} = c_{\circ} + c_1 h \ ; 
     \quad c_{\circ} + c_1 = 1 \ , 
\end{equation}
where $h$ is the cosine of the angle between the line of sight and the 
normal vector of the surface element of the photosphere. 
The Sun is known to follow the linear law in the optical bands \cite{gray}.  
We had adopted $c_1 = 0.6$ as a generic value for main sequence stars
in the calculation of the detection probability of microlensing
earth mass planets \cite{emplanet}.   
However, the linear models are suspected to be 
inadequate especially for hot stars,  and non-linear terms have been added in the 
functional form of the analytic models.  Recently, D\'iaz-Cordov\'es, Claret 
and Gim\'enez studied LTE (local thermodynamic equilibrium) model atmospheres 
(with standard solar compositions) 
calculated by Kurucz (1991) in search for suitable analytic limb-darkening models 
\cite{limb-sqrt,limb-linear,limb-lineartwo} that fit the numerical results.    
The limb-darkening models in their work can be written in the following form.
\begin{equation}
  {I(h)\over I_{\circ}} = \sum c_n h^n \ ; \quad  \sum c_n = 1 : 
       \quad n = 0, 0.5, 1, 2 
\label{eqProfile}
\end{equation} 
A homogeneous model is given by $c_n = 0; \ n \ne 0$,  a linear model is given 
by  $c_n = 0; \ n \ne 0, 1$,  and a non-linear model refers to a combination 
of the linear terms and one of the non-linear terms  ($c_2 \ne 0$: square model;
$c_{0.5} \ne 0$: square-root model).  The larger the index $n$, the more darkened
in the limb is the radial luminosity profile given by $h^n$.  $I_{\circ}$ is the
center luminosity (at $h=1$), and the condition $\sum c_n = 1$ ensures 
that the linear combination on the RHS of the equation (\ref{eqProfile}) 
has value $1$ at $h=1$.   

In a binary lens, a source star outside the caustic curve produces 
three normal images and a source star inside the caustic 
curve produces extra two bright images in addition to the three normal images.
At a caustic crossing, the surface of the star is divided into two regions 
by the caustic line moving at a constant velocity, and only the part of 
the star that falls inside the caustic produces two extra images. 
These partial images are connected 
at the {\it critical curve} (of the lens equation), 
and this criticality is what underwrites microlensing as a potential tool for
measuring stellar luminosity profiles.
\ 1)~ The magnification of these two images is governed by the  generic critical
behavior of lensing,  and this generic behavior simplifies 
the interpretation of the lensing.  
Thus, the light curve of the two partial images bears  direct 
measurements of the integrated luminosity profile of the source star (measured 
in time through the moving caustic line).  
\ 2)~  The critical curve is the loci of the images of divergent magnification 
of  point sources. Thus, the two partial images joined across the critical curve 
are highly magnified and dominate the behavior of the total light curve.
The three normal images are full images, insensitive to the luminosity
profile of the source star, and their brightness varies relatively slowly in time 
(unless the crossing is very near a cusp). 
     
The nomenclature {\it critical curve} derives from the fact that the lens 
equation is {\it critical}, or, {\it stationary} on the curve.   
One of the eigenvalues of the Jacobian matrix of the lens equation vanishes 
on the critical curve, and the lens equation is quadratic in the eigendirection 
of the vanishing eigenvalue.   As a consequence, the amplification of 
the images in the neighborhood of the critical curve is 
proportional to the inverse square root of the distance of the source to 
the caustic.   The proportionality constant that determines the overall 
peak amplification of the light curve of the two extra images is 
given by the inverse square root of the derivative of the Jacobian determinant
in the {\it critical direction} (the eigendirection of the  vanishing eigenvalue).   
General discussions of this critical behavior of lensing can be found in
the literature  ({\it e.g.},  \citeN{monograph_sef}).   However, 
we include a brief derivation  of  the critical behavior of 
the class of binary lenses  not only because the class is directly relevant
to  experiments, but also because its simplicity  allows  easy access to 
concrete understanding of the critical behavior.  

If the complex variables $z$ and $\omega$ denote an image position and its 
source position respectively, the binary lens equation is given as follows
\cite{bourassa,emplanet,98blg35}. 
\begin{equation}
 \omega = z - R_E^2\left({\epsilon_1\over \bar z - x_1} 
               + {\epsilon_2\over \bar z - x_2} \right) \ ; \qquad
       \epsilon_1 + \epsilon_2 = 1 
\label{eqBilens}
\end{equation}
where $\epsilon_1$ and $\epsilon_2$ are the fractional masses of the  lenses 
located at $x_1$ and $x_2$ respectively.  
The lens equation is intrinsically a relation between angular position variables,
and the lens plane (the complex plane parameterized by $z$ or $\omega$) can be 
put anywhere along the line of sight.  
The lens plane is a linear plane normal to the line of sight and
is equivalent to the angular space because scattering (deflection) angles involved
in microlensing are small ($<< 1$).  
If the lens plane is considered to pass through the center of the mass of the lenses, 
the Einstein ring radius of the total  mass $M$ of the binary lenses is given 
by  $R_E = \sqrt{4GMD}$ where $D$ is the reduced distance: 
~$D^{-1} = D_1^{-1} + D_2^{-2}$, and  $D_1$ and $D_2$ are the distances
along the line of sight (or in the radial direction) of the observer and
the source star from the lensing system respectively.  (It is conventional
that ``$R_E$" is reserved for the Einstein ring radius given by $\sqrt{4GMD}$.)   
The position variables can be rescaled by $R_E$ so that the unit distance scale
of the lens plane is set by the Einstein ring radius: ~$R_E = 1$.   (In
the following, all the position variables are measured in units of $R_E$. 
However, we may replace $1$ by $R_E$ when we find it useful for clarity.) 
The so-called image plane is the lens plane parameterized by the image position
variable (here $z$) and the source plane is the lens plane parameterized by the
source position variable (here $\omega$).

The linear (differential) behavior of the lens equation is given by
\begin{equation}
 {d\omega \choose d\bar\omega} 
   = \pmatrix{1  & \bar\kappa \cr
              \kappa &  1     \cr} {dz \choose d\bar z}  
   \equiv  \ {\cal J}\ {dz \choose d\bar z}  \ ,  
\label{eqJmatrix}
\end{equation}
where \  $z_i \equiv z - x_i : \ i = 1,2$, and  
\begin{equation}
 \kappa = {\epsilon_1\over z_1^2} + {\epsilon_2\over z_2^2}  \ 
        \equiv \  |\kappa|~ e^{2i\varphi} \ .   
\end{equation}
The eigenvalues of the Jacobian matrix ${\cal J}$  are
\begin{equation}
 \lambda_{\pm} = 1 \pm |\kappa| \ , 
\end{equation}
and the corresponding eigenvectors are
\begin{equation}
 e_+ \equiv {e^{-i\varphi} \choose e^{i\varphi}} \ , \quad
 e_- \equiv {i~e^{-i\varphi} \choose -i~e^{i\varphi}}  \ . 
\end{equation}   
Where $|\kappa| =1$, the eigenvalue $\lambda_-$ vanishes, and the Jacobian
matrix ${\cal J}$ behaves as a projection operator.  Thus, the curve defined
by $|\kappa| =1$ is called the critical curve.  Obviously, the Jacobian determinant 
$J = 1 - |\kappa|^2$ vanishes on the critical curve.  If $dz$  is a displacement from 
a point $z$ on the critical curve,  $d\omega$ is a non-vanishing displacement vector
parallel to the upper component of the eigenvector $e_+$ with eigenvalue 
$\lambda_+ = 2$.  Let  $E_+$ and $E_-$ denote the upper components of the eigenvectors
(the lower components are their complex conjugates), {\it i.e.},
$E_+ =  e^{-i\varphi}$ and  $E_- = i e^{-i\varphi}$.  Then, 
$dz \equiv  dz_+ E_+ + dz_- E_-$  where the linear coefficients $dz_\pm$ are real, 
and,   
\begin{equation}
   d\omega = \lambda_+ dz_+ E_+ = 2~dz_+ E_+ \ \equiv \delta\omega_{1+} E_+ \ .
\label{eqDomega}
\end{equation}
Because of this projection effect, the corresponding area element 
(in the $\omega$-plane) vanishes, 
$i/2 ~d\omega\wedge d\bar\omega= 0$, and the magnification of its image diverges:
$|J|^{-1} = dz\wedge d\bar z / d\omega\wedge d\bar\omega = \infty$.   
Now,  the caustic curve is, by definition, a curve on the $\omega$-plane
which the critical curve is mapped onto under the lens equation.  Thus,  
if $dz$ is an arbitrary displacement tangent to the critical curve, 
the induced displacement $d\omega$ in the $\omega$-plane is tangent to 
the caustic curve.  Since the induced displacement $d\omega$ is always in the
direction of $E_+$ as we have seen in equation (\ref{eqDomega}),   $E_+$ is 
tangent to the caustic curve everywhere;  or, the critical direction $E_-$ 
is normal to the caustic curve.     Therefore, when the source star approaches 
the caustic curve, the distance to the caustic $\delta$ can not be expressed as
a linear function of the image displacement from the critical curve $dz$.   
   If $\delta\omega_2$ is the second order contribution,  
\begin{equation}
  \delta\omega_2 =  ~{1\over 2}(-\bar\partial J)~(dz_+ - i~ dz_-)^2  \ .  
\end{equation}
If  $\delta\omega_2 = \delta\omega_{2+} E_+ + \delta\omega_{2-} E_-$, 
\begin{equation}
 \delta\omega_{2-} = ~{1\over 4} \left(\partial_+ J~ (2 dz_+ dz_-)
                        - \partial_- J (dz_+^2 - dz_-^2)\right) \ ,  
\label{eqQuadOne}
\end{equation} 
where $\partial_+ J$ and $\partial_- J$ are the derivatives of the Jacobian
determinant $J$ in the eigendirections.
\begin{equation}
 dJ = dz_+ \partial_+ J + dz_- \partial_- J  \ .
\label{eqDJ}
\end{equation}
Since $J=0$ on the critical curve, the Jacobian determinant near the critical
curve is simply $J = dJ$ in the linear approximation.   Now, we need to solve 
the lens equation to find $dz_+$ and $dz_-$ for a given displacement 
$\delta\omega$.   If we consider only the leading order terms,
\begin{equation}
 \delta\omega_+ = \delta\omega_{1+} \ , \quad
 \delta\omega_- = \delta\omega_{2-} \ , 
\end{equation}
then, $dz_+  =  \delta\omega_+/2$, and $dz_-$ is the solution of the following 
quadratic equation.  
\begin{equation}
 \delta\omega_- = {1\over 4} \left(\partial_- J~dz_-^2 
                + \partial_+ J~ \delta\omega_+  dz_-
                - \partial_- J {\delta\omega_+^2\over 4}\right) 
\label{eqQuad}
\end{equation}
Solutions exist where the discriminant is non-negative.   
\begin{equation}
  4~\delta\omega_- ~\partial_- J 
    + |\partial J|^2 ~{\delta\omega_+^2} \ge 0  \ . 
\label{eqDm}
\end{equation} 
The equality holds on the critical curve ($dJ = 0$)
as one can verify from equations (\ref{eqDJ}) and (\ref{eqQuad}).
The Jacobian determinant
can be calculated for the two solutions using (\ref{eqDJ}). 
\begin{equation}
  J = dJ = \pm\sqrt{4\partial_- J~\delta\omega_- 
                   + |\partial J|^2 \delta\omega_+^2} 
\end{equation} 
Now, let's note from the equality in (\ref{eqDm}) that the caustic curve is 
quadratic: \ $\delta\omega_- \propto  \delta\omega_+^2$ ~(if $\partial_- J \ne 0$).  
Therefore, the eigenvectors, $E_+$ and $E_-$, rotate (``frame rotation") 
as they move along the caustic curve.  
In this lowest approximation,  however, we can ignore the rotation, 
and the displacement of the point $(\delta\omega_+, \delta\omega_-)$ from the caustic
curve can be approximated by 
\begin{equation}
  \Delta\omega_- \equiv \delta\omega_- - \delta\omega_- (J =0) 
    \  = \delta\omega_- 
     + {|\partial J|^2 \over \partial_- J}{\delta\omega_+^2\over 4} \ .  
\end{equation}
The distance is $\delta = | \Delta\omega_-|$, and the Jacobian determinant takes a 
simple form.   
\begin{equation}
  J (\delta\omega_+, \delta\omega_-) 
   = \pm\sqrt{4\partial_- J~ \Delta\omega_-}  \ .
\end{equation} 
The two images have opposite parities, and their magnifications 
$A_{\pm}$ have a simple relation to the distance of the source to 
the caustic $\delta$. 
\begin{equation}
 A_{\pm} = {1 \over \sqrt{4|\partial_- J|}}{1\over \sqrt{\delta}} \ ; 
    \quad \delta \equiv |\Delta\omega_-| 
\label{eqCausAmp}
\end{equation}
The condition for the existence of the two extra images (\ref{eqDm}) can be 
rewritten elegantly as follows. 
\begin{equation}
 \partial_- J ~\Delta\omega_- > 0    
\label{eqInside}
\end{equation} 
Since the eigenvector $E_-$ is normal to the caustic curve,  if a source 
position with $\Delta\omega_- > 0$ is in one side of the caustic, then 
a source position  with $\Delta\omega_- < 0$ is in the other side of the caustic.  
So,  the inequality  (\ref{eqInside}) tells us  that  
a source in one side of the caustic curve (determined by the
direction vector $\partial_- J~E_-$)  generates two extra  images, 
but a source in the other side of the caustic does not.  Thus, a source   
star produces two partial images at a caustic crossing.  
In lensing where the wave property of the photon beam can be ignored, the
caustic curve is a geometric curve  with zero width, and  the source star 
is resolved by the caustic as a form of two partial images that change in 
time. (The side of the caustic where the two extra images are produced is
the inside of the caustic.)

The  inverse square-root dependence
of the amplification $A_\pm$ on the distance $\delta$ of a source position 
from the caustic curve in equation (\ref{eqCausAmp}) doesn't hold  if the 
second order contribution vanishes, $\delta\omega_2 = 0$.  The source
displacement from the caustic $\delta\omega_2$ induced from the image 
displacement $dz$ (from the critical curve) in equation (\ref{eqDm}) vanishes, 
when $\bar\partial J = 0$.   
\begin{equation}
 \bar\partial J = \bar\partial (1 - |\kappa|^2) 
     = -\kappa \bar\partial\bar\kappa
\end{equation} 
So, there are two cases where $\partial J$ vanishes: \ $\kappa= 0 $ or 
$\partial \kappa = 0$.  The condition  $\kappa = 0$ defines limit points where $J=1$
not the critical curve where $J=0$.  Therefore,  $\delta\omega_2$  vanishes only if
$\partial\kappa = 0$ which is the condition for bifurcation of $J=$ constant curves
\cite{binary}.   There are two cases where the critical curve has bifurcation points.  
If we let $\ell \equiv |x_1 - x_2|$ be the separation between the lens elements,
the critical curves with bifurcation points satisfy special relations between
the lens parameters $\ell$ and $\epsilon_2$ 
\cite{binary}.
The critical curve has one bifurcation point on the lens axis 
if $\ell = \ell_+$. 
\begin{equation}
    \ell_+ \ = \ \left({\root 3 \of {\epsilon_1}}
                   +{\root 3 \of {\epsilon_2}}\right)^{3\over 2}  \ \ ;
      \qquad 1  \ \le \ \ell_+ \ < \ 2  
\end{equation}
The critical curve has two bifurcation points off the lens axis 
if $\ell = \ell_-$.
\begin{equation}
    \ell_- \ = \ \left({\root 3 \of {\epsilon_1}}
                   +{\root 3 \of {\epsilon_2}}\right)^{-{3\over 4}} \ \ ;
       \qquad {1 \over \sqrt{2}} \ \le \ \ell_- \ < \ 1 
\end{equation}
Since the lens equation is continuous (actually differentiable) in the neighborhood
of the critical curve (the lens equation is singular only at the lens positions 
where  $J = -\infty$),  the caustic curve bifurcates where the critical curve 
bifurcates.   At the bifurcation points (four-prong vertex) of the critical 
curve, the lens equation is stationary along the critical curve.   
Therefore, the caustic curve forms a (doubly degenerate) cusp where it bifurcates.
If $\partial_- J = 0$, the critical curve is tangent to $E_-$, and the lens equation
is stationary with respect to a displacement along the critical curve.  Therefore,
the caustic curve forms a cusp where $\partial_- J = 0$.       On the bifurcation
points, both components $\partial_- J$ and $\partial_+ J$ vanish because 
the Jacobian determinant $J$ is stationary in every direction: $\partial J = 0$. 
Thus, if we consider only the line caustic crossings away from cusps, 
the quadratic behavior of the lens equation in the direction of $E_-$ remains valid, 
and the amplification of the two extra images across the 
critical curve can be discussed using the expression in equation (\ref{eqCausAmp}).   
The caustic curve has 8 cusps if ~$\ell > \ell_+$, 
6 cusps if ~$\ell_- < \ell < \ell_+$, and 10 cusps if ~$\ell < \ell_-$. 
 
\begin{figure}
\plotone{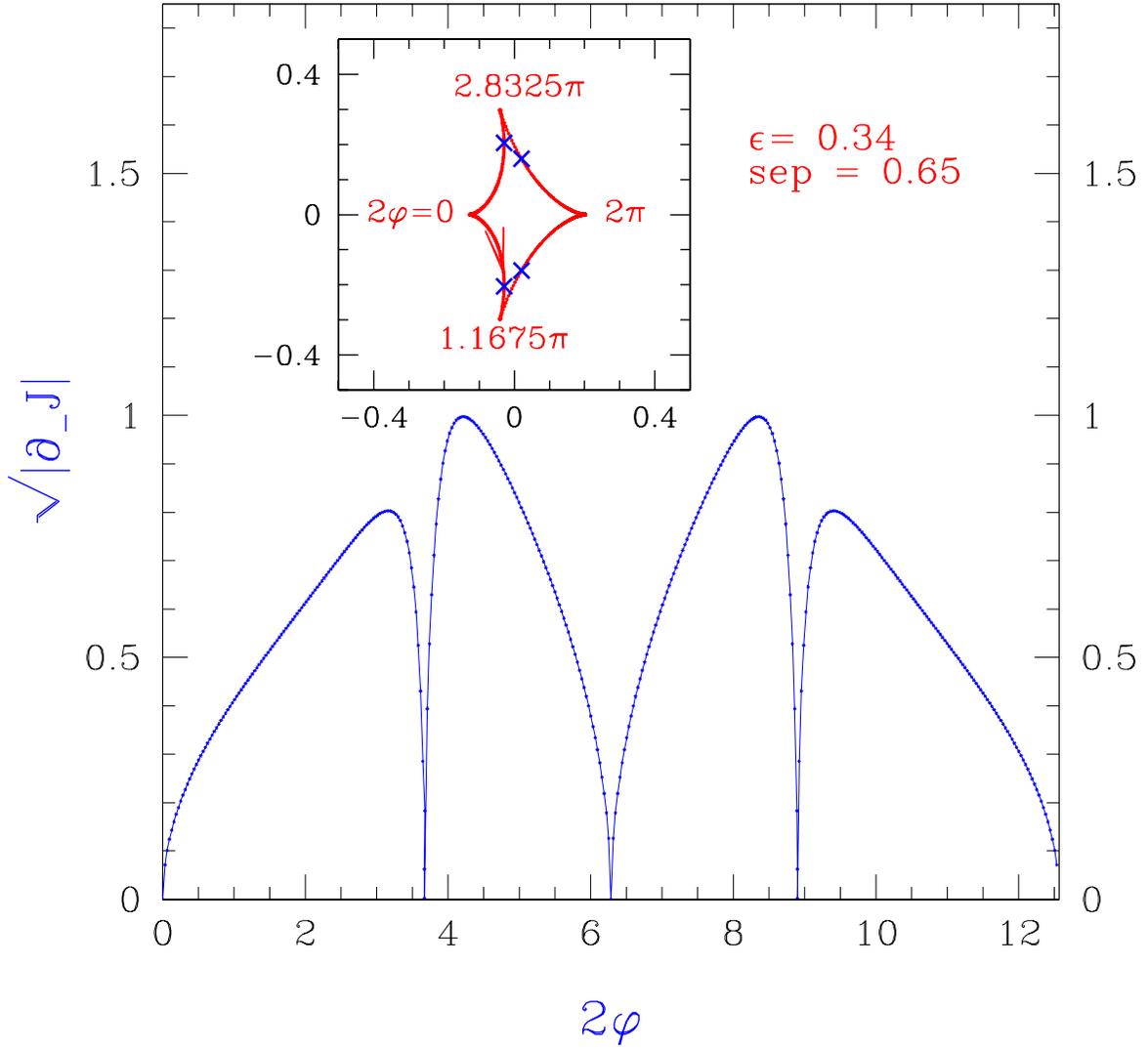}
\figcaption{\label{figcrit}
$\sqrt{|\partial_- J|}$ as a function of the phase angle of 
$\kappa = e^{i2\varphi}$:  The central caustic has four cusps, and 
$|\partial_- J|$ has four zeros.  Note that the total change of the phase 
angle along the caustic curve with four cusps is $4\pi$. (It is $2\pi$ for
a triangular caustic curve and $8\pi$ for a hexagonal caustic curve.   The 
total sum of the phase angle change along all the caustic loops of a binary
lens is  $8\pi$ which  is a topological invariant related to  the number
of point masses that is two for a binary lens.)  
}

\end{figure}

The amplification coefficient $|\partial_- J|^{-1/2}$ increases 
toward a cusp ($\partial_- J = 0$) where it diverges.  
Figure \ref{figcrit} shows the "central caustic" of a binary lens with
the fractional minor mass $\epsilon = \epsilon_2 = 0.34$ and the separation 
$\ell \equiv |x_1-x_2| = 0. 65$.  
The "central caustic" (always across the lens axis, and here contains the center of 
mass inside)  has four cusps,  and $\sqrt{|\partial_- J|}$ has four zeros at 
$2\varphi = 0, 1.1675\pi, 2\pi$, and $2.8325\pi$ where $2\varphi$ is the 
phase angle of ~$\kappa = e^{i2\varphi}$ (in the Riemann sheet).  The four 
blue crosses on the caustic curve mark the loci of the local maxima of 
$\sqrt{|\partial_- J|}$.  Since the amplification of the ``two extra images" 
is proportional to it, $|\partial_- J|^{-1/2}$ can be considered as  the 
``strength"  of the caustic curve.   (As we will discuss later,
the actual amplification of the two images also depends on the relative
size of the source star with respect to the Einstein ring radius, which
may be considered as a scale factor.)   Note that 
the local minima of $|\partial_- J|^{-1/2}$  are  very close to the
off-axis cusps.  We may loosely refer to it as that the cusps on the lens
axis are stronger than the ones off the lens axis.  In the same token, we
can say that the cusp at $\varphi = 0$ (near the dominant mass 
$\epsilon_1 = 0.66$) is stronger than the cusp at $\varphi = \pi$. 
In the case of MACHO-98-SMC-1,  the second
caustic crossing was closer to a cusp on the lens axis ($2\varphi = 0$)
than the first caustic to its nearest cusp ($2\varphi = 2\pi$), and
the caustic crossing amplification peak for the second caustic 
crossing is bigger than the first (see figure 1 and 2 in Rhie et~al 1999).
This lens ($\epsilon = 0.34$ and $\ell=0.65$) also has two triangular 
caustic curves off the lens axis (and reflection symmetric with respect 
to the lens axis -- real line here), 
and each has ``topological charge" $1/2$.  The eigenvectors $E_\pm$ rotate
by $\pi$ around each triangular caustic curve (or ``trioid").  The eigenvectors
rotate by $2\pi$ around the central caustic ``quadroid" in figure \ref{figcrit}
which has ``topological charge" $1$.    These off-axis caustics are ``weaker"
than the caustics which cross the lens axis.  For example, the maximum of 
$\sqrt{|\partial_- J|}$ on these ``trioids" (closed curve that is smooth except 
at three cusps) $\approx 6.93$ which is about 7 times larger than the maximum on the 
``central caustic" (``quadroid": closed curve that is smooth except at four cusps).

If $\partial_- J = 0$, equations (\ref{eqQuadOne}) or (\ref{eqQuad}) becomes 
linear in $dz_-$, and one obtains a linear solution whose amplification depends
linearly on the inverse of the distance of the source to the caustic: \ 
 $A_{\rm cusp} \propto 1/\delta$.  When the source exits the caustic through 
a cusp, this image crosses the critical curve.  This is the third image that 
diverges at a cusp crossing.   The other two image solutions can be  obtained
by including the third order terms in the expansion of $\delta\omega_{2-}$
in (\ref{eqQuadOne}).  If the source is sufficiently small compared to the
caustic curve,  the probability of crossing the caustic through a cusp is 
small. Out of about ten caustic crossing binary lensing events discovered
so far \cite{becker}, MACHO-97-BLG-28 was the only event where the crossing was
through a cusp \cite{97blg28-planet}.  The source star of the event 
MACHO-97-BLG-28 was a giant star.  Most of the lensed stars in microlensing 
experiments are main sequence stars.    
Here, we consider caustic crossings away from cusps only.  
 
If the source is sufficiently small, the curvature of the caustic
curve can also be ignored, and we can assume that the luminosity variation
of the light curve over the caustic crossing is largely determined by 
the critical behavior of the lensing amplification and the luminosity
profile of the source star.   If we let the line caustic be along the 
$y$-direction,  the lensing amplification is independent of $y$,  and we only need
to know the ``1-d luminosity profile" of the source in the normal direction,
say $x$-direction,  which is obtained by integrating the 2-d profile along 
the $y$-direction.   It is worth emphasizing that  the luminosity profile we 
probe directly in a line-caustic crossing microlensing event is 
the 1-d luminosity profile of the source star.  (The ``irregularities" of 
the stars such as star spots are not addressed here.   The 1-d luminosity
profile will depend on the position of the spots with respect to the direction
of the caustic line.)

\begin{figure}
\plotone{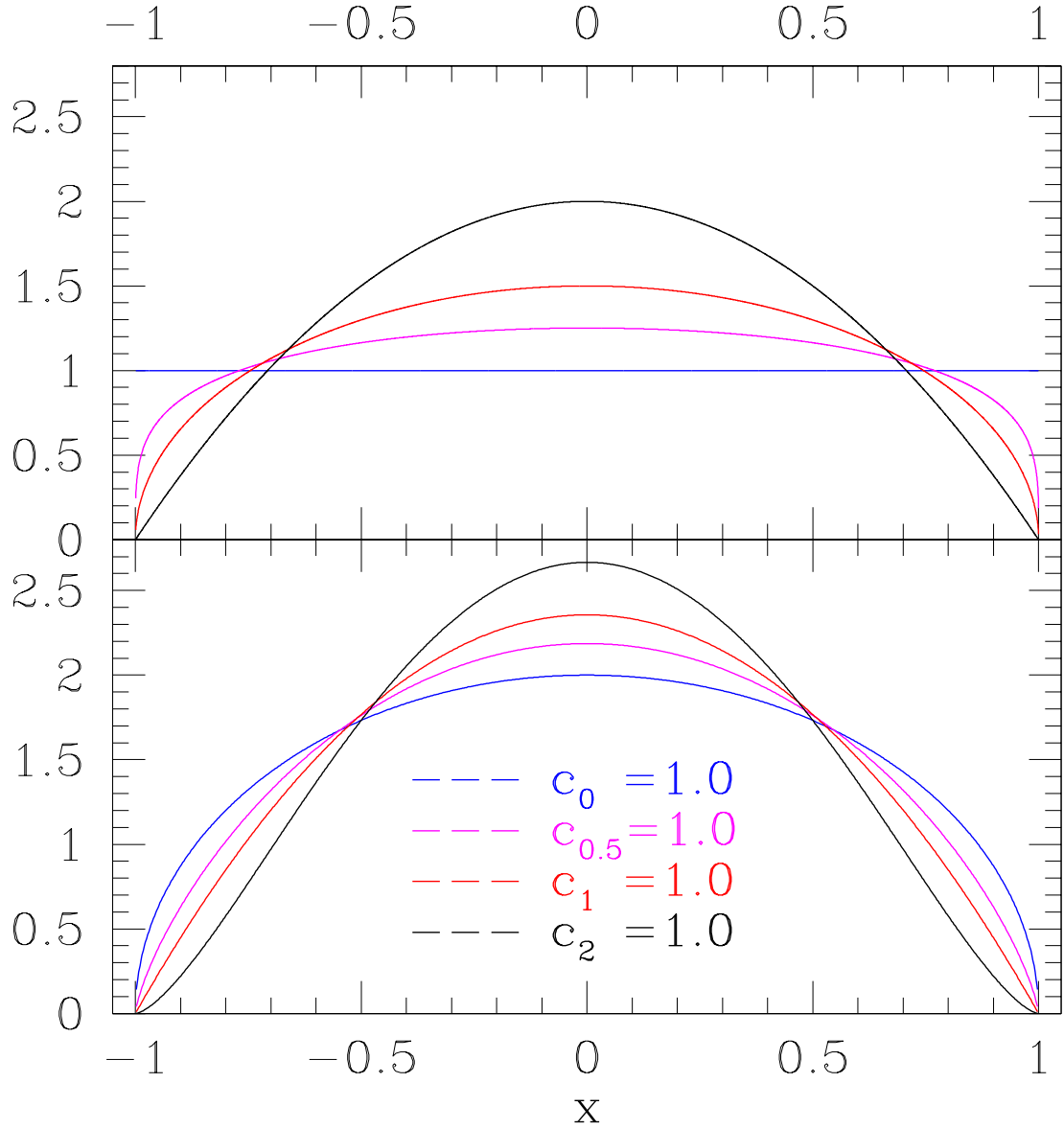}
\figcaption{\label{fig-lumprofiles}
The luminosity profile functions across the diameter of the stellar disk 
(upper panel) and 1-d luminosity profile functions (lower panel)
for $n=0, 0.5, 1, 2$:  The total flux emitted perpendicular to the disk 
(or toward an observer) has been normalized to $\pi$.  In a line caustic crossing 
microlensing, one can effectively scan the 1-d luminosity profile in time.
}
\end{figure}

If the size of the star is $r_\ast$ (in units of Einstein ring radius), 
the luminosity profile  in equation (\ref{eqProfile}) can be rewritten 
in terms of the radial 
coordinate $r$ of the stellar disk using $ h^2 = 1 - r^2/r_\ast^2 $\~. \
We let $r^2 = r_\ast^2~(x^2 + y^2)$ such that ~$x$ and $y$ are the Cartesian 
coordinates of the unit stellar disk, and  define  a dimensionless 
1-d luminosity function ~$f(x) = \sum_n c_n f_n(x)$ based on the radial 
luminosity profile in equation (\ref{eqProfile}).
\begin{equation}
   f_n(x) \ \equiv \ \int h^n dy = \int_0^{\sqrt{1 - x^2}} 
     \left( 1 - r^2/ r_\ast^2 \right)^{n\over 2} dy 
        = 2 b_n \left(1 - x^2\right)^{n+1\over 2} \ ,
\label{eqLuminosity}
\end{equation}
where 
$b_0 = 1, \ b_{0.5} = 0.87362649, \ b_1 = \pi/4$, \ and $b_2 = 2/3$.   
$$ 
 b_n \equiv \int_0^1 \left(1 - t^2 \right)^{n\over 2} dt 
$$ 
The integrated flux is given by a byproduct of ${b_n}'s$.
\begin{equation}
  F_n \ \equiv \ \int h^n ~dx dy = \int f_n ~dx = 4 b_n b_{n+1}
\end{equation}
We find $b_{1.5} = 0.71849148$~ and \ $b_3 = 3\pi/16$,  and
$$
 F_0 = \pi~ ; \quad
 F_{0.5} = 2.51077224  \approx {4\pi\over 5}~ ; \quad
 F_1 =  {2\pi\over 3}~ ; \quad
 F_2 =  {\pi\over 2}
$$
Figure \ref{fig-lumprofiles} shows normalized radial luminosity profile 
~$h_n \pi/F_n$ and 1-d luminosity profile  $f_n \pi/F_n$.  We have chosen 
the total flux to be $\pi$ for easy recognition of the peak values of the 
radial luminosity profile in the plot: \  the peak values 
are $1, ~\approx 5/4, ~3/2$, and $2$ for $n = 0, 0.5, 1$, and $2$  
respectively.  For the 1-d luminosity profiles, the peak values are  given
by $\pi/2b_{n+1}$: \ $2.00, ~2.19, ~2.36$, and ~$2.67$.     
Because of the integration in $y$,  the limb-darkening effect is more
accentuated in the 1-d luminosity profile compared to
the radial luminosity profile of the stellar disk.

The amplification profile of the two extra images at a caustic crossing can 
be calculated using the luminosity profile $f(x)$ and the analytic form of the
point source amplification function $A_\pm$ in equation (\ref{eqCausAmp}).
\begin{equation}
    {\int fA ~dx \over \int f ~dx } \ \equiv G \quad ; 
     \qquad A \equiv A_+ + A_-
\end{equation}   
Since the point source amplification function $A$ vanishes in one side 
of the caustic line, we may specify the segment of the one 
dimensional star (as defined by the 1-d luminosity function) that is inside the 
caustic, say, as given by an interval  $[r_\ast a, ~r_\ast]$~ where $ (a < 1)$.     
Then, $A$ may be written as a function of $x$ accordingly.
\begin{equation}
   A = {1\over \sqrt{|\partial_-J|}} {1\over \sqrt{r_\ast}}
       {1\over \sqrt{x - a}} \ \equiv \ A_c A_\ast A_x 
\end{equation}
The first factor $A_c$ is determined by where on the caustic curve the crossing 
occurs.  Figure \ref{figcrit} shows $A_c^{-1}$ for the central caustic curve of 
a lens with $(\epsilon, ~\ell) = (0.34, ~0.65)$.  In this case, in a line caustic 
crossing, $A_c \sim 1$-$10$.  The other triangular caustics we mentioned above
but not shown here are ``weaker" and $A_c \sim 0.15$-$1$ for a line caustic 
crossing.    The second factor $A_\ast$ depends on the size of the star. 
\begin{equation}
   r_\ast = D_1 \alpha_\ast = {R_\ast \ D_1 \over D_1 + D_2}  
          = {\alpha_\ast\over \alpha_E} \ , 
\end{equation}  
where $\alpha_\ast$ and $R_\ast$ are angular (apparent) and the physical 
sizes of the star respectively, and $\alpha_E = R_E/D_1$ is the angular size 
of the Einstein ring radius.  Here, $\alpha_E = 1/D_1$ because we set $R_E=1$.
 If $\alpha_E \sim 1$~mas, and 
$\alpha_\ast \sim 1$ - $10\mu$as, then $A_\ast \sim 30$ - $10$.   

If $c_n =1$ for a given $n$, the amplification profile function $G$ is determined
by one function $f_n$,  and we may define $G_n \equiv G_{| c_n=1}$.  Then, for 
an arbitrary luminosity profile, $G$ is written as a linear combination of $G_n$'s.    
\begin{equation}
   G(a) = {\sum_n \tilde c_n~ G_n(a)\over  \sum_n \tilde c_n~} :
    \qquad \tilde c_n = c_n~F_n~ ; \quad \sum_n c_n = 1
\end{equation}
The appearance of the ``new" coefficients $\tilde c_n$ is due to that
the directly measurable quantity in lensing is the amplification of the images
where  the  normalization is the total flux while the coefficients $c_n$'s
have been defined based on the luminosity profile where the more relevant 
unit quantity is the peak luminosity.   Since we need to be able to compare 
the microlensing measurements  with theoretical calculations as well 
as measurements from other methods, it should be useful to consider the linear
coefficients $c_n$ of the analytic luminosity profile functions as the basic
parameters and feel free to redefine linear coefficients based on them as  
the needs arise. 

Since the amplification factors $A_c$ and $A_\ast$ contribute to  the
amplification of the two extra images as overall scale factors, we define
$g_n(a)$ that depends only on $A_x(a)$ such that $G_n(a) = A_c A_\ast g_n(a)$.
Then, $G(a) = A_c A_\ast g(a)$  where the shape function 
$g(a)$ is given as follows.
\begin{equation}
    g(a) = {\sum_n \tilde c_n~ g_n(a)\over  \sum_n \tilde c_n~}  
\end{equation}
In an observed lightcurve, the ``time" $a$ in units of the (relative) stellar radius 
$r_\ast$ should be replaced by the actual time variable $t$.  If the caustic crossing 
occurs at $t = t_{cc}$ (when the center of the star crosses the caustic line: $a=0$),
$t_\ast$ is the stellar radius crossing time, and $\iota$ is the incident angle of 
the source trajectory (the small angle between $\pm E_-$ and the source trajectory),
\begin{equation}
 a = {(t-t_{cc})~\cos\iota \over t_\ast} \ .  
\end{equation}
So, the time axis of an observed lightcurve depends on the stretch factor
\begin{equation}
 {t_\ast\over \cos\iota} = {\alpha_\ast\over \mu \cos\iota}
 = {r_\ast R_E\over v_\perp \cos\iota}
 = { R\ast\over v_\perp \cos\iota}~ { D_1\over D_1+D_2} \ , 
\end{equation}  
where $\mu = |\vec \mu|$ is the relative proper motion 
of the lens with respect to the
source star (as seen from the observer) and $v_\perp = |\vec v_\perp|$ is the 
relative transverse linear velocity of the lens.  

\begin{figure}
\plotone{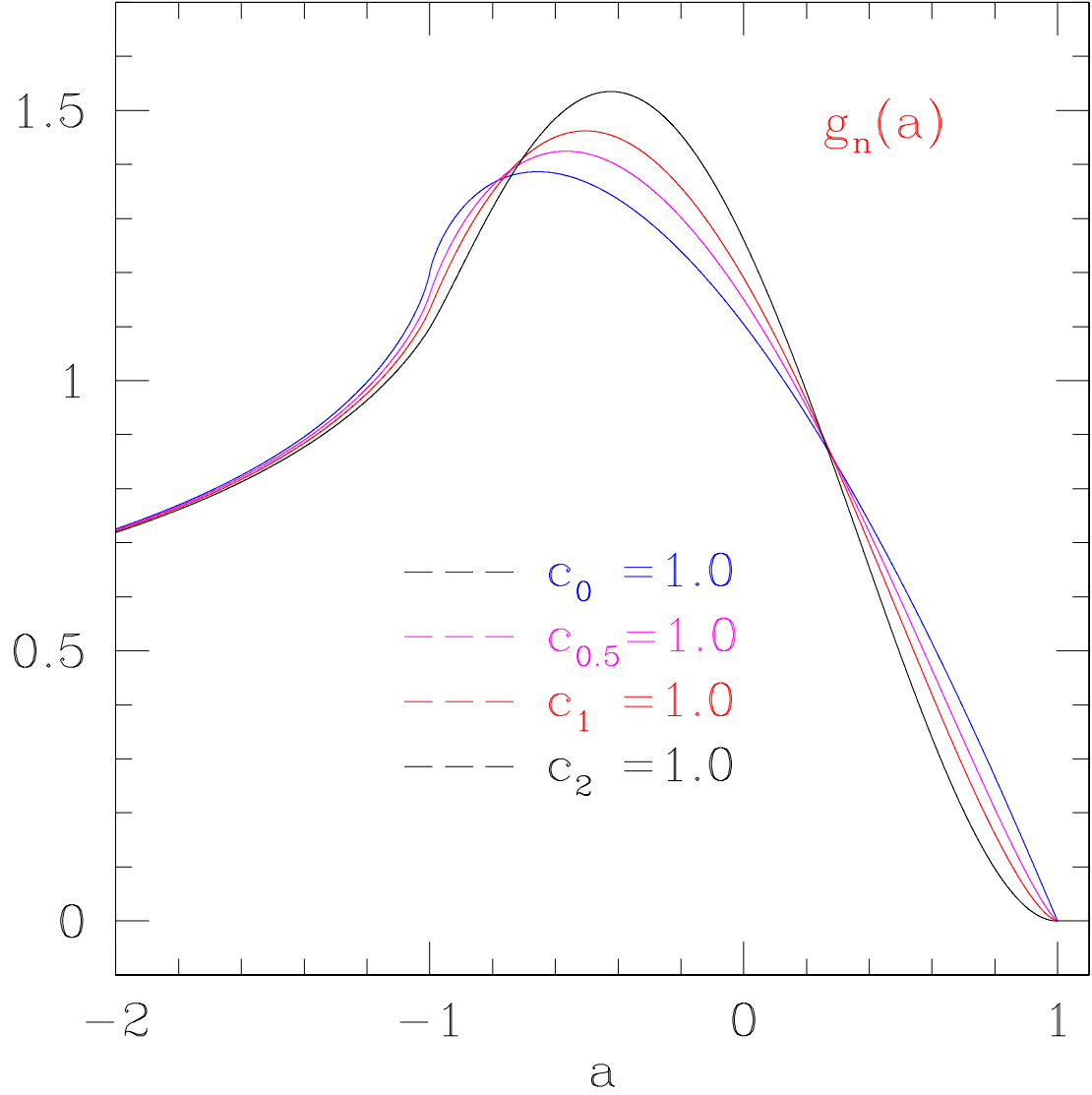}
\figcaption{\label{fig-amprofiles}
Light curves of the two extra images of stars with luminosity profiles
given by $f_n: n = 0, 0.5, 1, 2$.  $a$ is the position of the center of 
the stellar disk with respect to the caustic line     
in units of the radius of the star $r_ast$. 
The caustic line touches  the limb of the star from inside a caustic 
at $a = -1$ and leaves the caustic at $a = 1$.  The luminosity
profiles are indistinguishable when the star inside the
caustic line is about one diameter away from the caustic ($a = -2$).  
The peak crossing times are sensitive to the luminosity profiles.
The peak crossing times are at $a =-0.656, ~-0.565, ~-0.503$, and
$-0.423$ for $n=0, ~0.5, ~1$ and $2$.
}
\end{figure}

\begin{figure}
\plotone{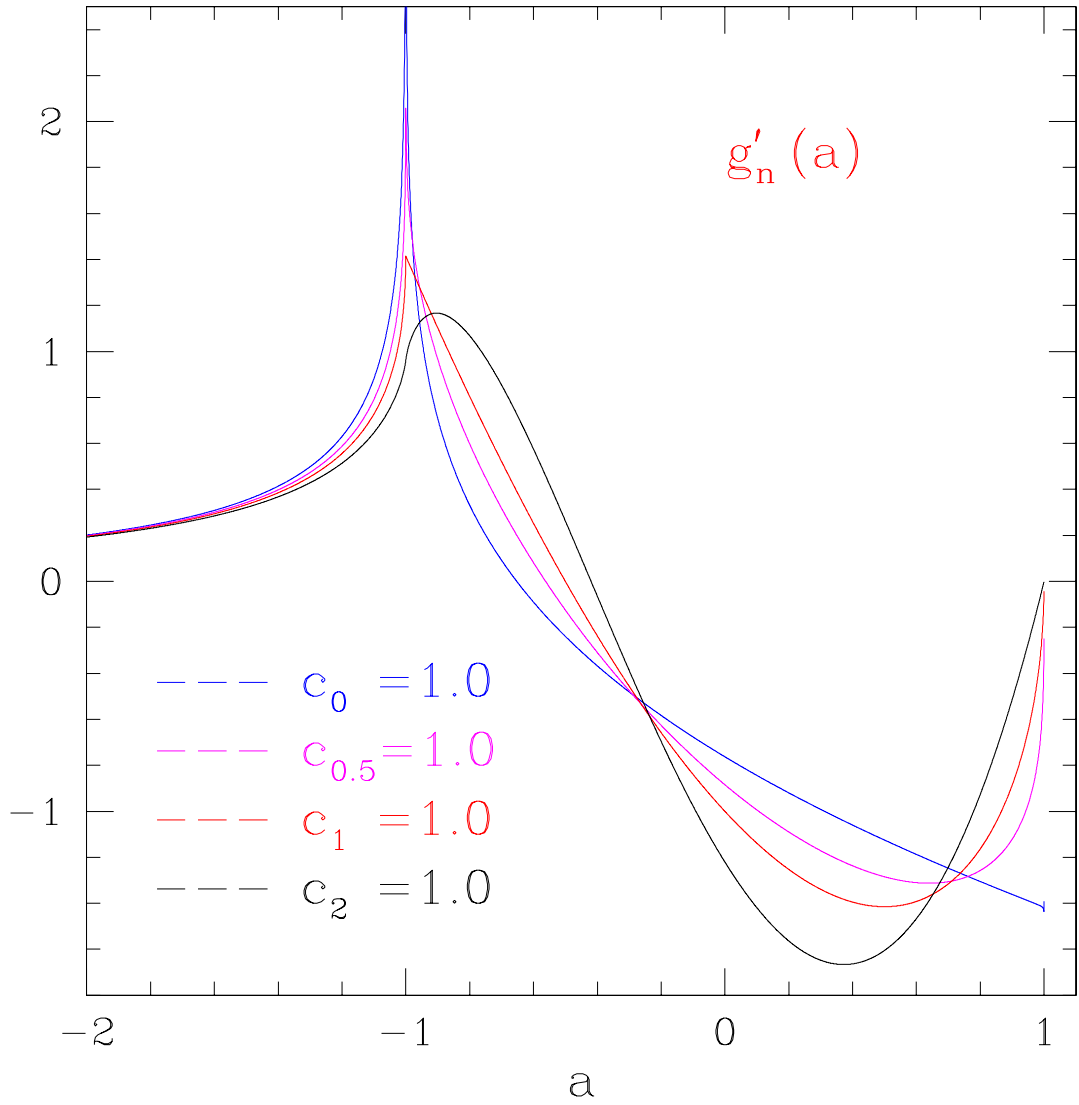}
\figcaption{\label{fig-slope}
The derivatives $g_n^{\prime}(a)$: \ at $a=-1$, the 
slope of the shape functions $g_n(a)$ change singularly. 
 At the end-point ($a = 1$), the slope
is discontinuous for $n=0$, and continuous for the others.  
}
\end{figure}

If the caustic line is at $x = 0$, and the center of the star is at $x = a < 0$, 
then the amplification profile of the partial images ($a \ge -1$) is given as 
follows.
\begin{equation}
   g_n(a) \equiv  {\int f_n A_x ~dx \over \int f_n ~dx }   
           = {1\over 2 b_{n+1}} 
        \int_a^1 {(1-x^2)^{n+1\over 2}\over\sqrt{x-a}}~dx 
\label{eqGna}
\end{equation}
For the two full extra images of the star when it is completely inside the 
caustic curve ($a < -1$), the integration range should be extended to  $[-1, 1]$.    
The functions $g_n(a)$~ in figure \ref{fig-amprofiles} and \ref{fig-slope}
show some of the characteristic effects of the luminosity profiles at a line
caustic crossing:  
\begin{enumerate}
 \item  The lightcurve becomes practically insensitive to the 
luminosity profile when the  star is  about two stellar radii away from the caustic
line  ($a = -2$).  
\item  In the immediate neighborhood of $a=-1$, the lightcurve
rises rapidly and the curvature of the light curve changes from concave upward
to convex.  Thus, given a well-sampled lightcurve, the caustic entry ``time"  
$a=-1$, where the limb of the star touches the caustic line producing two images
joined at a critical point, can be read off from the lightcurve with a relatively 
small error.  Past $a=-1$, the images become partial, and the loss of the
flux due to the partial imaging causes the slope of the lightcurve to decrease
leading to peak turn-over and rapid decline of the lightcurve.    
\item  The peaks 
are relatively broad, and the practically universal intersection points of the 
falling curves at $a = a_f \approx 0.28$ offer a natural distinction between 
the peak ($a < a_f$) and the tail ($a_f < a < 1$).  With increasing $n$~
(the surface luminosity is more concentrated toward the center of the star with
larger $n$), the rising curve of the peak becomes less steeper and the falling
curve becomes more steeper.   (The slope of $g_{\circ}$ diverges at $a=-1$.)
\item  The peak
crossing time where the lightcurve has the maximum amplification 
and  the peak amplification value are  sensitive to the luminosity profile.
The peak crossing times are  $a =-0.656, ~-0.565, ~-0.503$, and 
$-0.423$ for $n=0, ~0.5, ~1$ and $2$ respectively.  (The peak crossing time 
of a point source is $a =0$.)   The varying peak crossing times look a bit more
impressive  in unnormalized lightcurves \cite{buxton}.  
\item  The tails ($a_f < a < 1$) are dominated by the ``linear behavior"
except very near the end point ($a =1$).  Thus, given a well-sample lightcurve,
the caustic exit ``time" $a=1$, where the limb of the star completely exits
the caustic and the two extra images disappear, can be more or less read off  
from the lightcurve.   The slope of the lightcurve can change abruptly 
($n =0$) or smoothly ($n = 0.5, 1, 2$)) at the end-point (see below).   
Near the end-point, the observed lightcurve is dominated by three normal images 
whose magnification behavior outside the caustic curve crucially depends on the 
proximity to a nearby cusp.  (A cusp dominates the magnification behavior of the 
source plane around it more like the point caustic of a single and may be 
considered as a point caustic with directionality.)
If the flux of the normal images is about 10\% of the peak flux,  the shape
function may not be representative of the behavior of the observed lightcurve
when $a \gsim 0.8$.  In the case of MACHO-98-SMC-1, the flux of the normal images
was about $7\%$ of the peak flux for the second caustic crossing and about $10\%$
for the first caustic crossing \cite{98smc1-mps}.   In the event MACHO-97-BLG-41, 
it was about $16\%$ for the central caustic crossing and about $30\%$ for the 
planetary caustic crossing \cite{97blg41,97blg41-planet}.     
\item The caustic crossing time,  $a=0$,  marks a most featureless part of 
the lightcurve.   
\end{enumerate}
 
In order to examine the the end-point behavior of the shape function $g_n(a)$, 
we rewrite the expression in (\ref{eqGna}) with $\xi = 1 -x$. 
\begin{equation}
 g_n(a) = {1\over 2 b_{n+1}} 
   \int_0^{1-a} {(2\xi-\xi^2)^{n+1\over 2}\over\sqrt{1-a-\xi}}~d\xi 
\end{equation}
For $a \approx 1$, the higher order term $\xi^2$ in the numerator can be
ignored, and $g_n(a)$ scales in $(1-a)$ as follows.
\begin{equation}
 g_n(a) \ \propto \  (1-a)^{n+2\over 2} 
\end{equation}
where the proportional constant is given as follows. 
\begin{equation}
  {1\over 2 b_{n+1}} 
   \int_0^1 {(2\eta)^{n+1\over 2}\over\sqrt{1-\eta}} ~d\eta
\end{equation}
Therefore, the derivative of $g_n$ vanishes at $a=1$ unless $n=0$, which can
also be seen in figure \ref{fig-slope}. 
For a homogeneous model,  the shape function $g_{\circ}$ 
has a finite slope at the end-point: \  $g_{\circ}^{\prime}(1) = -\sqrt{2}$.  
This has an interesting ramification that every
lightcurve of line caustic crossing should exhibit an abrupt change of the
slope at the end-point.  That is because  the luminosity profile of any star 
has (or is believed to have) a sharp edge, or, equivalently, a nonvanishing 
$c_{\circ}$.    

\begin{figure}
\plotone{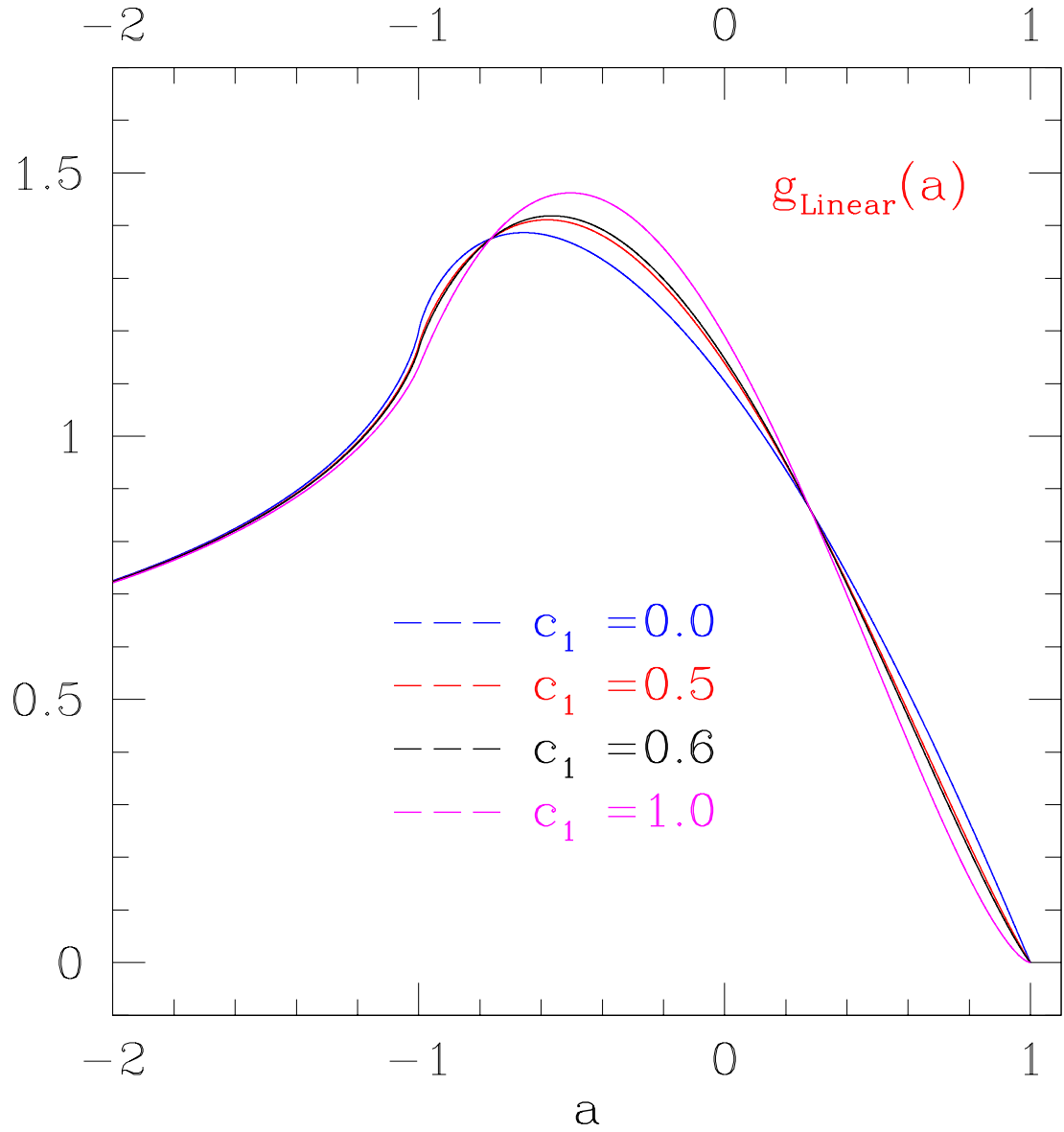}
\figcaption{\label{fig-linear}
Light curves of the linear limb-darkening models:  $g_{\circ}$
and $g_1$ determine the intersection points $a = -0.76$ and $0.28$, and the shape
functions of linear models distributed in between $g_{\circ}$ and $g_1$.  The typical
linear limb-darkening parameters $c_1=0.5$ and $c_1=0.6$ are indistinguishable in 
this plot.  
}
\end{figure}

\begin{figure}
\plotone{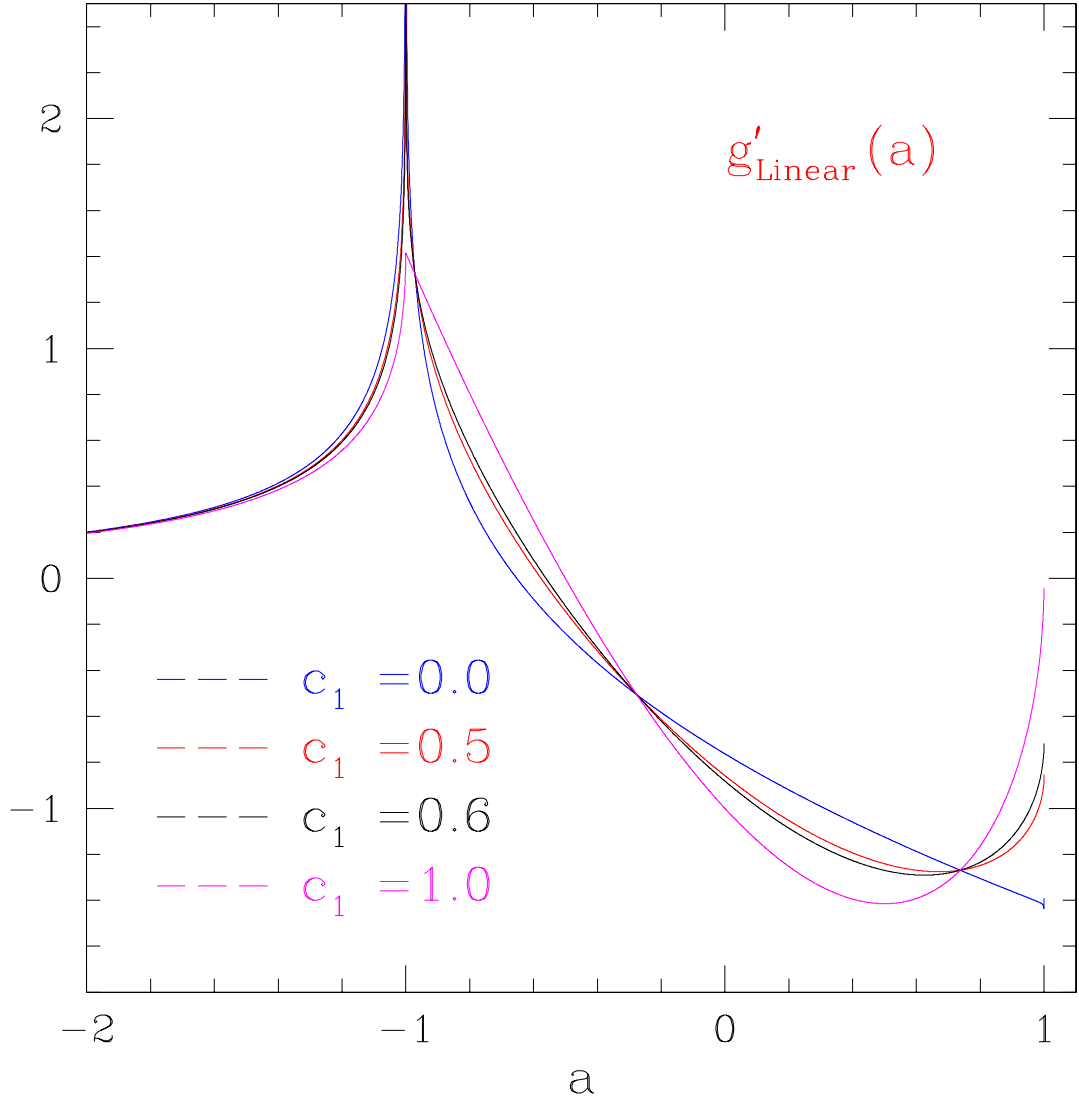}
\figcaption{\label{fig-slope-linear}
The linear models with $c_1 = 0.5$ and $0.6$ show relatively constant slopes 
between $a \approx 0.3$-$0.8$.  Toward the end-point, the slopes change rapidly 
from $g^{\prime} \approx -1.2$ to $-\sqrt{2} c_{\circ}$.     
}
\end{figure}

\begin{figure}
\plotone{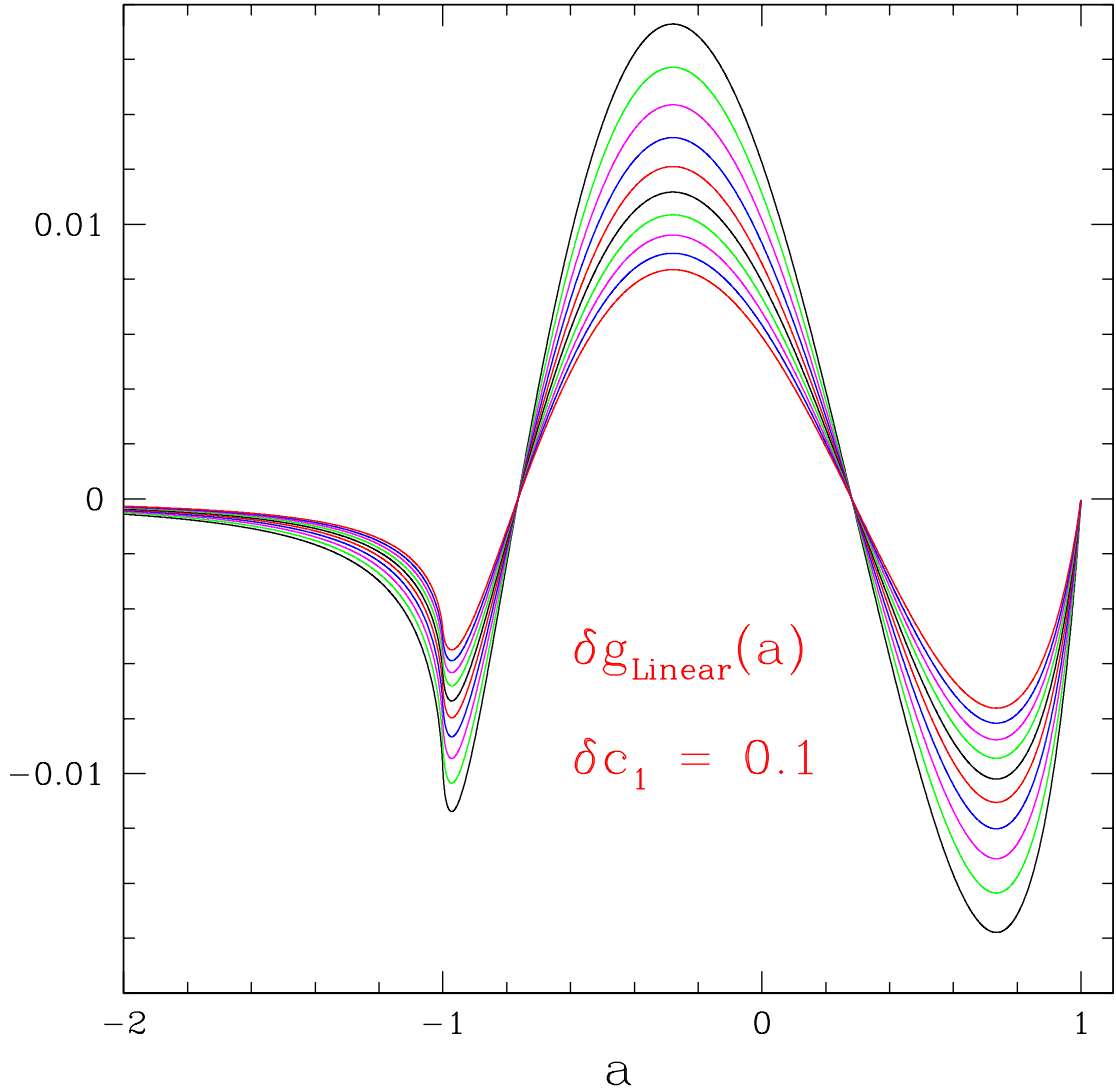}
\figcaption{\label{fig-delc}
Difference lightcurves for linear models with $c_1$ values that
differ by $\delta c_1 = 0.1$: \ $\delta g_{Linear}(a: c_1 + \delta c_1)
 - \delta g_{Linear}(a: c_1): ~\delta c_1 = 0.1$ for $c_1 = 0.0, ... , 0.9$.
The difference lightcurve with the largest amplitude is for $c_1=0.9$. 
}
\end{figure}

\begin{figure}
\plotone{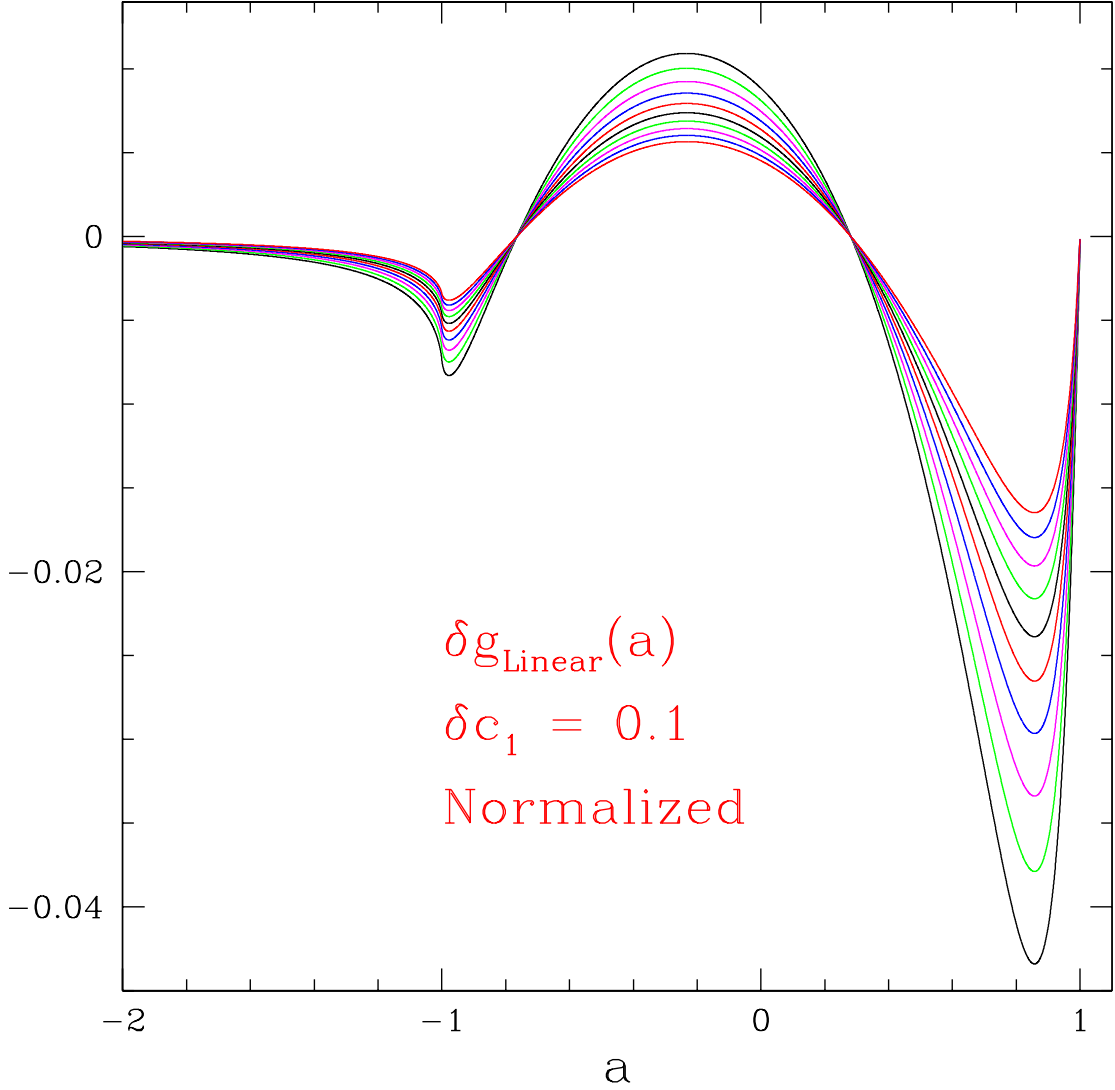}
\figcaption{\label{fig-delc-norm}
Normalized difference lightcurves for linear models with $c_1$ values that
differ by $\delta c_1 = 0.1$: \ the difference lightcurves in figure 7 
have been divided by $g(a: c_1+0.5 \delta c_1) + 0.2$.  The baseline value
$0.2$ we chose here as an example amounts to $\approx 12\%$ of the peak 
amplification.    In order to determine the linear limb-darkening parameter 
$c_1$ with $\delta c_1 = 0.1$,  photometric accuracy of  $0.3$-$0.8~\%$  
is needed.
}
\end{figure}

In order to gauge how well one can determine the luminosity profiles,
we examine the linear models.   
\begin{equation}
  g_{Linear}(a) = {\tilde c_{\circ} g_{\circ}(a) + \tilde c_1 g_1(a) 
                    \over \tilde c_{\circ} + \tilde c_1} \ ; \qquad
               c_{\circ} + c_1 = 1
\end{equation} 
Figure \ref{fig-linear} shows that the shape functions for $c_1=0.5$ and 
$c_1=0.6$ are almost indistinguishable in this plot.  The intersection points
are at $a= -0.76$ and $a= 0.28$, and it will be necessary to sample
the lightcurves in all three regions defined by the intersection points.  
The derivatives of the shape functions are shown in figure \ref{fig-slope-linear}. 
The curvature changes at $a=-1$ marks the entry time, and it will be relatively
easy to reconstruct the correct time in a fitting.   The linear models with 
$c_1=0.5$ and $c_1=0.6$ show relatively constant slopes between $a=0.3$ and $0.8$
or so.  The opposite trends of the slopes of $g_{\circ}$ and $g_1$ make the slope
of the  shape function of a typical main sequence star with a medium value of
the linear limb-darkening parameter (say, $c_1=0.3-0.7$) more or less constant. 
This ``linear behavior" of the lightcurve in the tail was measured for the first
time in the event MACHO-98-SMC-1 \cite{98smc1-eros}.  

Figure \ref{fig-delc} shows the differences 
$\delta g_{Linear} (a: c_1) \equiv 
g_{Linear} (a: c_1+\delta c_1) - g_{Linear} (a: c_1):
\ \delta c_1=0.1 $ for $c_1 = 0, 0.1, .... , 0.9$.   
In this figure, the three regions
divided by the intersection points are manifest.   The larger the value of $c_1$, 
the higher the peak value at $a = -0.28$ and the larger amplitude of deviation.  
Thus, the smaller the value of $c_1$, the more demanding is the photometric 
accuracy and precision.   In order to see the level of photometric accuracy we need, 
we normalize the difference curve $\delta g_{Linear} (a: c_1)$ by 
$g_{Linear} (a: c_1+ 0.5 \delta c_1) + 0.2$ as an example.  
Figure \ref{fig-delc-norm} shows the normalized difference curves.   In order 
to determine the linear limb-darkening
parameter $c_1$, one needs to be able to reconstruct the shape function, and this
requires  that all three regions defined by the intersections points $a=-0.76$
and $a=0.28$  be sampled fairly.  The differences in tails look more 
discriminatory.  However,  the tails are more susceptible to the effects of the 
normal images and blending, and should be interpreted properly case by case.   
In an observed lightcurve, the determination of the scale factors $A_c$ and 
$A_\ast$ can be compromised if one overinterprets the ``linear behavior" 
of the tails.   

In summary, we have analyzed the magnification behavior of the two extra images 
that appear and disappear in a line caustic crossing  binary microlensing. 
We have identified the shape function of the lightcurve of the two extra images
and examined its dependence on the luminosity profile of the lensed star.   
We also discussed the multiplicative factors that determines the overall amplification
and the ones that scale the time axis in an observed lightcurve.  
We suggest that it is important to sample the lightcurve fairly 
in all three regions that are defined by the 
intersection points.  The span of the lightcurve that is affected by the luminosity
profile is about $3 r_\ast$~ ($a= [-2, 1]$).  In the case of linear models, we
estimate from figure \ref{fig-delc-norm} 
that the photometric accuracy of $0.3$-$0.8~\%$ is required
to be able to determine the linear limb-darkening parameter $c_1$ with precision 
of $\delta c_1 = 0.1$.   It is similar to the estimation made for the eclipsing 
binary method by Popper (1984).   In lensing, the signal lies in magnification
while it is in absorption in the case of eclipsing binaries.  Thus, microlensing 
will have advantage for fainter stars.

\acknowledgements
\section*{Acknowledgments}

This research has been supported in part by the NASA Origins
program (NAG5-4573), the National Science Foundation (AST96-19575), and
by a Research Innovation Award from the Research Corporation.


\def\ref@jnl#1{{\rm#1}}
\def\aj{\ref@jnl{AJ}}
\def\apj{\ref@jnl{ApJ}}
\def\apjl{\ref@jnl{ApJ}}
\def\apjs{\ref@jnl{ApJS}}
\def\aap{\ref@jnl{A\&A}}
\def\aapr{\ref@jnl{A\&A~Rev.}}
\def\aaps{\ref@jnl{A\&AS}}
\def\mnras{\ref@jnl{MNRAS}}
\def\prl{\ref@jnl{Phys.~Rev.~Lett.}}
\def\pasp{\ref@jnl{PASP}}
\def\nat{\ref@jnl{Nature}}
\def\iauc{\ref@jnl{IAU~Circ.}}
\def\aplett{\ref@jnl{Astrophys.~Lett.}}
\def\annrev{\ref@jnl{Ann.~Rev.~Astron.~and Astroph.}}

\clearpage


%
%

%
%

\end{document}